\documentclass[aps,floatfix]{revtex4}
\usepackage{graphics}
\usepackage{epsfig}
\usepackage{subfigure}
\usepackage{amsmath,amsfonts,amssymb,graphicx,times}
\begin{document}

\title{Personal Recommendation via Modified Collaborative Filtering}

\author{Run-Ran Liu$^{a}$}
\author{Chun-Xiao Jia$^{a}$}
\author{Tao Zhou$^{a,b}$}
\email{zhutou@ustc.edu}
\author{Duo Sun$^{a}$}
\author{Bing-Hong Wang$^{a,c}$}

\affiliation{$^{a}$ Department of Modern Physics and Nonlinear
Science Center, University of Science and Technology of China, Hefei
Anhui, 230026, PR China}

\affiliation{$^{b}$ Department of Physics, University of Fribourg,
Chemin du Muse 3, CH-1700 Fribourg, Switzerland}

\affiliation{$^{c}$ Institute of Complex Adaptive System, Shanghai
Academy of System Science, Shanghai, P. R. China}

\date{\today}

\begin{abstract}
In this paper, we propose a novel method to compute the similarity
between congeneric nodes in bipartite networks. Different from the
standard cosine similarity, we take into account the influence of
node's degree. Substituting this new definition of similarity for
the standard cosine similarity, we propose a modified collaborative
filtering (MCF). Based on a benchmark database, we demonstrate the
great improvement of algorithmic accuracy for both user-based MCF
and object-based MCF.
\end{abstract}

\maketitle



\section{Introduction}
Recently, recommendation systems are attracting more and more
attentions, because it can help users to deal with information
overload, which is a great challenge in the modern society,
especially under the exponential growth of the Internet \cite{1} and
the World-Wide-Web \cite{2}. Recommendation algorithm has been used
to recommend books and CDs at Amazon.com, movies at named
Netflix.com, and news at VERSIFI Technologies (formerly
AdaptiveInfo.com) \cite{3}. The simplest algorithm we can use in
these systems is global ranking method (GRM) \cite{15}, which sorts
all the objects in the descending order of degree and recommends
those with highest degrees. GRM is not a personal algorithm and its
accuracy is not very high because it does not take into account the
personal preferences. Accordingly, various kinds of personal
recommendation algorithms are proposed, for example, the
collaborative filtering (CF) \cite{4,5}, the content-based methods
\cite{6,7}, the spectral analysis \cite{8, renjie}, the principle
component analysis \cite{9}, the diffusion approach
\cite{15,16,18,ljg}, and so on. However, the current generation of
recommendation systems still requires further improvements to make
recommendation methods more effective \cite{3}. For example, the
content analysis is practical only if the items have well-defined
attributes and those attributes can be extracted automatically; for
some multimedia data, such as audio/video streams and graphical
images, the content analysis is hard to apply. The collaborative
filtering usually provides very bad predictions/recommendations to
the new users having very few collections. The spectral analysis has
high computational complexity thus infeasible to deal with huge-size
systems.

Thus far, the widest applied personal recommendation algorithm is CF
\cite{3,10}. The CF has two categories in general, one is user-based
(U-CF), which recommends the target user the objects collected by
the users sharing similar tastes; the other is object-based (O-CF),
which recommends those objects similar to the ones the target user
preferred in the past. In this paper, we introduce a modified
collaborative filtering (MCF), which can be implemented for both
object-based and user-based cases and achieve much higher accuracy
of recommendation.

\section{Method}
We assume that there is a recommendation system which consists of
$m$ users and $n$ objects, and each user has collected some objects.
The relationship between users and objects can be described by a
bipartite network. Bipartite network is a particular class of
networks \cite{15,17}, whose nodes are divided into two sets, and
connections among one set are not allowed. We use one set to
represent users, and the other represents objects: if an object
$o_{i}$ is collected by a user $u_{j}$, there is an edge between
$o_{i}$ and $u_{j}$, and the corresponding element $a_{ij}$ in the
adjacent matrix \emph{A} is set as 1, otherwise it is 0.

In U-CF, the predicted score $v_{ij}$ (to what extent $u_{j}$ likes
$o_{i}$), is given as :
\begin{equation}
v_{ij}=\sum^{m}_{l=1,l\neq{i}}s_{il}a_{jl},
\end{equation}
where $s_{il}$ denotes the similarity between $u_{i}$ and $u_{l}$.
For any user $u_{i}$, all $v_{ij}$ are ranked by values from high to
low, objects on the top and have not been collected by $u_{i}$ are
recommended.

How to determine the similarity between users? The most common
approach taken in previous works focuses on the so-called structural
equivalence. Two congeneric nodes (i.e. in the same set of a
bipartite network) are considered structurally equivalent if they
share many common neighbors. The number of common objects shared by
users $u_i$ and $u_j$ is
\begin{equation}
c_{ij}=\sum^{n}_{l=1}a_{li}a_{lj},
\end{equation}
which can be regarded as a rudimentary measure of $s_{il}$.
Generally, the similarity between $u_i$ and $u_j$ should be somewhat
relative to their degrees \cite{11}. There are at least three ways
previously proposed to measure similarity, as:
\begin{eqnarray}
s_{ij}=\frac{2c_{ij}}{k(u_i)+k(u_j)},&\\
s_{ij}=\frac{c_{ij}}{\sqrt{k(u_i)k(u_j)}},&\\
s_{ij}=\frac{c_{ij}}{min(k(u_i),k(u_j))}.
\end{eqnarray}
The Eq.(3) is called Sorensen's index of similarity (SI) \cite{12},
which was proposed by Sorensen in 1948; the Eq.(4), called the
cosine similarity, was proposed by Salton in 1983 and has a long
history of the study on citation networks \cite{11}; the Eq.(5) is
called Pearson correlation. Both the Eq.(4) and Eq.(5) are widely
used in recommendation systems \cite{3,15}.

A common blemish of Eqs. (3)-(5) is that they have not taken into
account the influence of object's degree, so the objects with
different degrees have the same contribution to the similarity. If
user $u_{i}$ and $u_{j}$ both have selected object $o_{l}$, that is
to say, they have a similar taste to the object $o_{l}$. Provided
that object $o_{l}$ is very popular (the degree of $o_{l}$ is very
large), this taste (the favor for $o_{l}$) is a very ordinary taste
and it does not means $u_{i}$ and $u_{j}$ are very similar.
Therefore, its contribution to $s_{ij}$ should be small. On the
other hand, provided that object $o_{l}$ is very unpopular (the
degree of $o_{l}$ is very small), this taste is a peculiar taste, so
its contribution to $s_{ij}$ should be large. In other words, it is
not very meaningful if two users both select a popular object, while
if a very unpopular object is simultaneously selected by two users,
there must be some common tastes shared by these two users.
Accordingly, the contribution of object $o_{l}$ to the similarity
$s_{ij}$ (if $u_i$ and $u_j$ both collected $o_l$) should be
negatively correlated with its degree $k(o_{l})$. We suppose the
object $o_{l}$'s contribution to $s_{ij}$ being inversely
proportional to $k^{\alpha}(o_{l})$, with $\alpha$ a freely tunable
parameter. The $s_{ij}$, consisted of all the contributions of
commonly collected objects, is measured by the cosine similarity as
shown in Eq. (4). Therefore, the proposed similarity reads:
\begin{equation}
s_{ij}=\frac{1}{\sqrt{k(u_i)k(u_j)}}\sum^{n}_{l=1}\frac{a_{li}a_{lj}}{k^{\alpha}(o_l)}.
\end{equation}
Note that, the influence of object's degree can also be embedded
into the other two forms, shown in Eq. (3) and Eq. (5), and the
corresponding algorithmic accuracies will be improved too. Here in
this paper, we only show the numerical results on cosine similarity
as a typical example.

\begin{figure}
\scalebox{0.8}[0.8]{\includegraphics{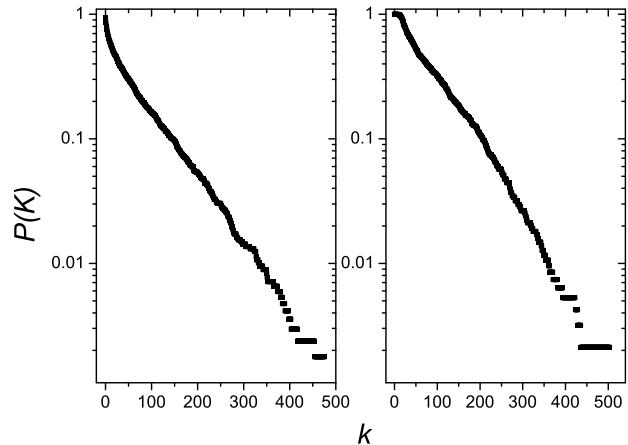}} \caption{The
degree distributions of users (left panel) and objects (right panel)
in linear-log plot, where $P(k)$ denotes the cumulative degree
distribution.}
\end{figure}

For any user-object pair $u_{i}$-$o_{j}$, if $u_{i}$ has not yet
collected $o_{j}$, the predicted score can be obtained by using Eq.
(1). Here we do not normalize Eq. (1), because it will not affect
the recommendation list, since for a given target user, we need sort
all her uncollected objects, and only the relative magnitude is
meaningful. Note that, if two objects have exactly the same score,
their order is randomly assigned. We call this method a modified
collaborative filtering (U-MCF), for it belongs to the framework of
U-CF.

\section{Numerical results}
Using a benchmark data set namely \emph{MovieLens} \cite{13}, we can
evaluate the accuracy of the current algorithm. The data consists of
1682 movies (objects) and 943 users. Actually, \emph{MovieLens} is a
rating system, where each user votes movies in five discrete ratings
1-5. Hence we applied a coarse-graining method used in Refs.
\cite{15,16}: A movie has been collected by a user if and only if
the giving rating is at least 3 (i.e. the user at least likes this
movie). The original data contains $10^{5}$ ratings, 85.25$\%$ of
which are $\geq 3$, thus the data after the coarse gaining contains
85250 user-object pairs. The current degree distributions of users
and objects were presented in Fig. 1. Clearly, the degree
distributions of both users and objects obey an exponential form. To
test the recommendation algorithms, the data set is randomly divided
into two parts: The training set contains 90$\%$ of the data, and
the remaining 10$\%$ of data constitutes the probe. Of course, we
can divided it in other proportions, for example, 80$\%$ \emph{vs}.
20$\%$, 70$\%$ \emph{vs}. 30$\%$, and so on. The training set is
treated as known information, while no information in probe set is
allowed to be used for prediction.

\begin{figure}
\scalebox{0.8}[0.8]{\includegraphics{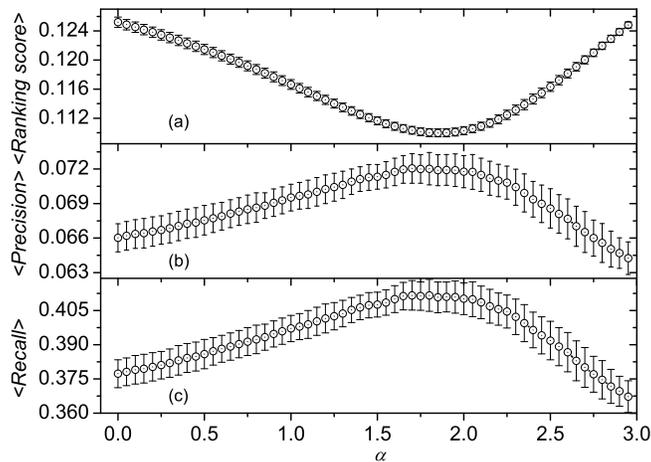}} \caption{The
effect of parameter $\alpha$ in U-MCF. The ranking score has its
minimal at about $\alpha$ = 1.85, at almost the same point, the
recall and precision achieve their maximums. Present results are
obtained by averaging over four independent 90$\%$ \emph{vs}. 10$\%$
divisions. The error bars denote the standard deviations. }
\end{figure}

A recommendation algorithm could provide each user a recommendation
list which contains all her/his uncollected objects. There are
several measures for evaluating the quality of these recommendation
lists generated by different algorithms. In this paper, we use
\emph{ranking score}, \emph{recall} and \emph{precision} to measure
the effectiveness of a given recommendation approach. Good overview
of these measures can be found in Ref  \cite{5}.

\emph{Ranking score}. For an arbitrary user $u_{i}$, if the relation
$u_{i}$-$o_{j}$ is in the probe set (according to the training set,
$o_{j}$ is an uncollected object for $u_{i}$), we measure the
position of $o_{j}$ in the ordered queue. For example, if there are
1000 uncollected movies for $u_{i}$, and $o_{j}$ is the 10th from
the top, we say the position of $o_{j}$ is the top 10/1000, denoted
by $r_{ij}$ = 0.01. Since the probe entries are actually collected
by users, a good algorithm is expected to give high recommendations
to them, thus leading to small $r$. Therefore, the mean value of the
position value $\langle{r}\rangle$ (called ranking score \cite{15}),
averaged over all the entries in the probe, can be used to evaluate
the algorithmic accuracy. The smaller the ranking score, the higher
the algorithmic accuracy, and vice verse. The definition of ranking
score here is slightly different from that of the Ref. \cite{15}. It
is because if a movie or user in the probe set has not yet appeared
in the training set, we automatically remove it from the probe and
the number of total movies was counted only for the ones appeared in
the the training set; while the Ref. \cite{15} takes into account
those movies only appeared in the probe via assigning zero score to
them. This slight difference in implementation does not affect the
conclusion.

\emph{Recall} is defined as the ratio of number of recommended
objects appeared in the probe to the total number of objects. The
larger recall corresponds to the better performance. Recall is also
called hitting rate in literature  \cite{15}.

\emph{Precision} is defined as the ratio of number of recommended
objects appeared in the probe to the total number of recommended
objects. The larger precision corresponds to the better performance.
Recall and precision depend on the length of recommendation list
\emph{L}, we set \emph{L} as 50 in our numerical experiment (in real
e-commerce systems, the length of recommendation list usually ranges
from 10 to 100 \cite{Schafer2001}), therefor the total number of
recommended objects is $mL$ = 47150.

Fig. 2 reports the algorithmic accuracy of U-MCF, which has a clear
optimal case around $\alpha$ = 1.85. Fig. 3 (a) reports the
distribution of all the position values, $r_{ij}$, which are sorted
from the top position ($r_{ij}$$\rightarrow$0) to the bottom
position ($r_{ij}$$\rightarrow$1). Fig. 3 (b) and (c) report the
recall and precision for different lengths of recommendation lists
respectively. Fig. 4 reports the algorithmic accuracies of the
standard case ($\alpha$ = 0) and the the optimal cases ($\alpha$ =
1.85) for different sizes of training sets. All these numerical
results strongly demonstrate that to depress the contribution of
common selected popular objects can further improve the algorithmic
accuracy.

\begin{figure}
\scalebox{0.8}[0.8]{\includegraphics{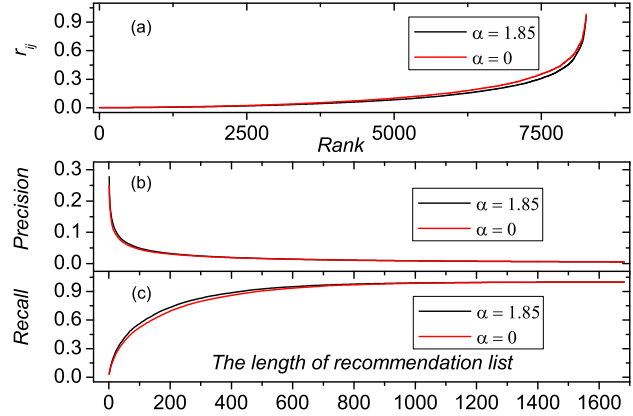}} \caption{(Color
online) (a): The predicted position of each entry in the probe
ranked in the ascending order. (b): The precision for different
lengths of recommendation lists. (c): The recall for different
lengths of recommendation lists.}
\end{figure}

\begin{figure}
\scalebox{0.8}[0.8]{\includegraphics{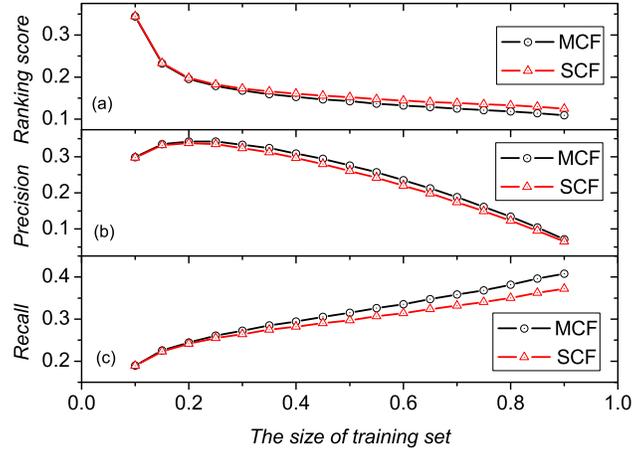}} \caption{(color
online) The standard CF (SCF) (i.e. $\alpha$ = 0 ) \emph{vs}. the
optimal case for different sizes of training sets.}
\end{figure}

\begin{figure}
\scalebox{0.8}[0.8]{\includegraphics{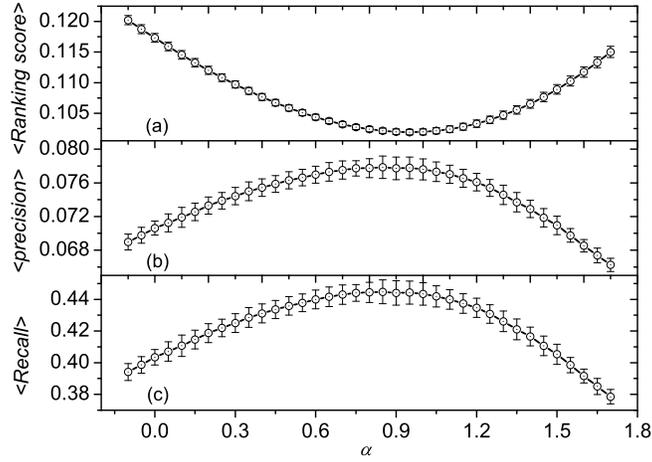}} \caption{The
effect of parameter $\alpha$ in O-MCF. The ranking score has its
minimal at about $\alpha$ = 0.95, at almost the same point, the
recall and precision achieve their maximums. Present results are
obtained by averaging over four independent 90$\%$ \emph{vs}. 10$\%$
divisions. The error bars denote the standard deviations.}
\end{figure}

Similar to the U-CF, the recommendation list can also be obtained by
object-based collaborative filtering (O-CF), that is to say, the
user will be recommended objects similar to the ones he/she
preferred in the past \cite{Sarwar2001}. By using the cosine
expression, the similarity between two objects, $o_i$ and $o_j$, can
be written as:
\begin{equation}
s_{ij}=\frac{1}{\sqrt{k(o_i)k(o_j)}}\sum^{m}_{l=1}a_{il}a_{jl}.
\end{equation}
The predicted score, to what extent $u_i$ likes $o_j$, is given as:
\begin{equation}
v_{ij}=\sum^{n}_{l=1,l\neq{i}}s_{jl}a_{li}.
\end{equation}
\begin{table}
\caption{Three measures for different algorithms with probe set
containing 10$\%$ data. For precision and recall, $L=50$. Present
results are obtained by averaging over four independent divisions.
The values corresponding to U-MCF and O-MCF are the optimal ones.}

\begin{tabular}{llll}
method &$<$Ranking score$>$&$<$Precision$>$&$<$Recall$>$\\
\hline
GRM &  0.1502  &  0.3077  &  0.0540  \\
O-CF &  0.1173  &  0.4035  &  0.0706  \\
U-CF &  0.1252  &  0.3773  &  0.0660  \\
O-MCF &  0.1019  &  0.4443  &  0.0777  \\
U-MCF &  0.1101  &  0.4108  &  0.0719  \\
\hline
\end{tabular}
\end{table}

Analogously, taking into account the influence of user degree, a
modified expression of object-object similarity reads:
\begin{equation}
s_{ij}=\frac{1}{\sqrt{k(o_i)k(o_j)}}\sum^{m}_{l=1}\frac{a_{il}a_{jl}}{k^{\alpha}(u_{l})},
\end{equation}
where $\alpha$ is a free parameter. The modified object-based
collaborative filtering (O-MCF for short) can be obtained by
combining Eq. (8) and Eq. (9). Fig. 5 reports the algorithmic
accuracy of O-MCF, which has a clear optimal case around
$\alpha=0.95$. Fig. 6 (a) reports the distribution of all the
position values, $r_{ij}$, which are sorted from the top position
($r_{ij}$$\rightarrow$0) to the bottom position
($r_{ij}$$\rightarrow$1), Fig. 6 (b) and (c) report the recall and
precision for different lengths of recommendation lists
respectively. Fig. 7 reports the algorithmic accuracies of the
standard case ($\alpha$ = 0) and the the optimal case ($\alpha$ =
0.95) for different sizes of training sets. All these results,
again, demonstrate that to depress the contribution of users with
high degrees to object-object similarity can further improve the
algorithmic accuracy of object-based method.
\begin{figure}
\scalebox{0.8}[0.8]{\includegraphics{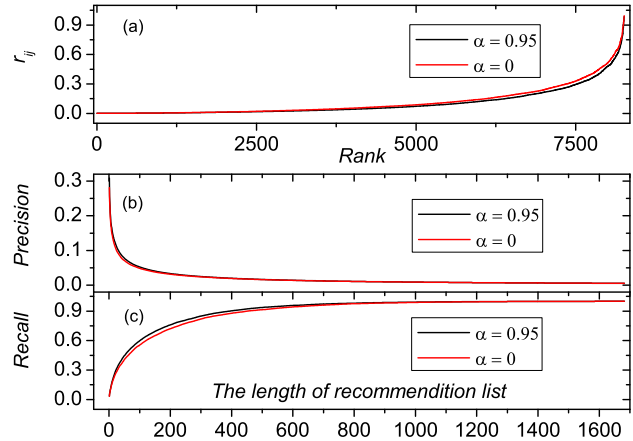}} \caption{(color
online) Similar to Fig.3. But for O-MCF.}
\end{figure}

\begin{figure}
\scalebox{0.8}[0.8]{\includegraphics{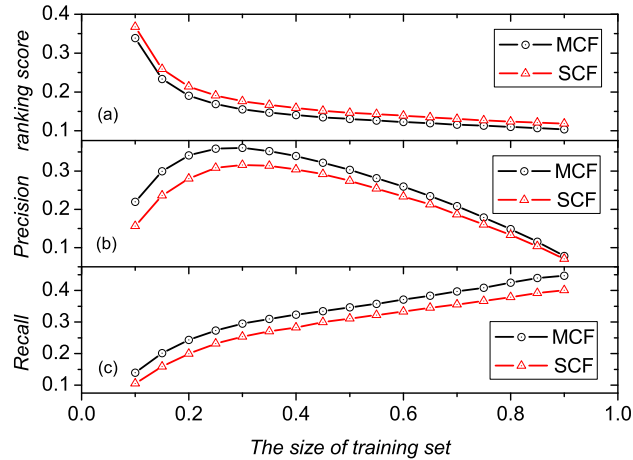}} \caption{(color
online) Similar to Fig.4. But for O-MCF.}
\end{figure}

\section{conclusion}
We compare the MCF, standard CF and GRM in Tab. I. Clearly, MCF is
the best method and GRM performs worst. Compared with the standard
CF, the modified object-based algorithm and the modified user-based
method improve the accuracy in different extent in three measures.
Ignoring the degree-degree correlation in user-object relations, the
algorithmic complexity of U-MCF is
$O(m^{2}\langle{k_{u}}\rangle+mn\langle{k_{o}}\rangle)$, the O-MCF
is $O(n^{2}\langle{k_{o}}\rangle+mn\langle{k_{u}}\rangle)$,
respectively. Here $\langle{k_{u}}\rangle$ and
$\langle{k_{o}}\rangle$ denote the average degree of users and
objects. Therefore, one can choose either O-MCF or U-MCF according
to the specific property of data source. For example, if the user
number is much larger than the object number (i.e. $m \gg n$), the
O-MCF runs much faster. On the contrary, if $n \gg m$, the U-MCF
runs faster. Furthermore, the remarkable improvement of algorithmic
accuracy also indicates that our definition of similarity is more
reasonable than the traditional one.

\section*{ACKNOWLEDGMENTS}
We acknowledge \emph{GroupLens Research Group} for providing us the
data set \emph{MovieLens}. This work is funded by the National Basic
Research Program of China (973 Program No.2006CB705500), the
National Natural Science Foundation of China (Grant Nos. 60744003,
10635040, 10532060 and 10472116), and the Specialized Research Fund
for the Doctoral Program of Higher Education of China. T.Z.
acknowledges the support from SBF (Switzerland) for financial
support through project C05.0148 (Physics of Risk), and the Swiss
National Science Foundation (205120-113842).

\end{document}